\documentclass[12pt, aps, prd, letterpaper, amsmath, amssymb, notitlepage, tightenlines, superscriptaddress, floatfix, nofootinbib, longbibliography]{revtex4-1}
\usepackage[T1]{fontenc}
\usepackage{graphicx}
\usepackage{amsmath}
\usepackage{amsfonts}
\usepackage[colorlinks=true, linkcolor=blue, urlcolor=blue, citecolor=blue, anchorcolor=blue]{hyperref}
\usepackage{multirow}

\newcommand{\alat}{\ensuremath{a_{\text{lat}}} }
\renewcommand{\d}{\ensuremath{\mathrm d} }
\newcommand{\cD}{\ensuremath{\mathcal D} }
\newcommand{\Emin}{\ensuremath{E_{\text{min}}} }
\newcommand{\Emax}{\ensuremath{E_{\text{max}}} }
\newcommand{\cN}{\ensuremath{\mathcal N} }
\newcommand{\Nj}{\ensuremath{N_{\text{j}}} }
\newcommand{\cO}{\ensuremath{\mathcal O} }
\newcommand{\de}{\ensuremath{\delta} }
\newcommand{\si}{\ensuremath{\sigma} }
\newcommand{\vev}[1]{\ensuremath{\left\langle #1 \right\rangle} }
\newcommand{\eq}[1]{Eq.~\ref{#1}}
\newcommand{\fig}[1]{Fig.~\ref{#1}}
\newcommand{\secref}[1]{Section~\ref{#1}}
\newcommand{\refcite}[1]{Ref.~\cite{#1}}

\begin{document}
\title{First-order bulk transitions in large-$N$ lattice Yang--Mills theories using the density of states}

\author{Felix Springer}\email{Felix.Springer@liverpool.ac.uk}
\affiliation{Department of Mathematical Sciences, University of Liverpool, Liverpool L69 7ZL, United Kingdom}

\author{Enrico Rinaldi}
\affiliation{Interdisciplinary Theoretical and Mathematical Sciences Program (iTHEMS), RIKEN, 2-1 Hirosawa, Wako, Saitama 351-0198, Japan}

\author{David Schaich}\email{david.schaich@liverpool.ac.uk}
\affiliation{Department of Mathematical Sciences, University of Liverpool, Liverpool L69 7ZL, United Kingdom}

\collaboration{for the Lattice Strong Dynamics (LSD) Collaboration}
\noaffiliation

\begin{abstract}
We use the Logarithmic Linear Relaxation (LLR) density of states algorithm to study the bulk phase transitions of pure-gauge SU($N$) lattice Yang--Mills theories with $4 \leq N \leq 8$.
This approach avoids super-critical slowing down at such transitions, which poses a problem for traditional importance sampling Monte-Carlo methods. We analyse the effect of different updating strategies within the LLR algorithm, different reconstruction techniques of the density of states and different lattice volumes. By comparing our results for the weakly first-order SU(5) bulk phase transition against those for the stronger transitions with $N \geq 6$, we demonstrate the advantages of the LLR method for analyses of strong transitions with large latent heat.
\end{abstract}

\maketitle

\section{Introduction}
\label{intro}
Many strongly interacting systems exhibit first-order phase transitions characterized by a non-zero latent heat --- a discontinuity in the energy density.
Such first-order transitions in the early universe would produce a stochastic background of gravitational waves, like the one recently observed in the nanohertz frequency range by several pulsar timing array collaborations~\cite{NANOGrav:2023gor, EPTA:2023fyk, Reardon:2023gzh, Xu:2023wog}.
Future space-based facilities including the LISA observatory~\cite{Caprini:2015zlo, Caprini:2019egz}, DECIGO~\cite{Kawamura:2020pcg} and AEDGE~\cite{AEDGE:2019nxb} will also search for such backgrounds at higher frequencies.
This has motivated renewed interest in non-perturbative analyses of first-order transitions in strongly coupled gauge theories, for which first-principles lattice field theory calculations are a crucial tool --- see \refcite{Aarts:2023vsf} for a recent review.

QCD-like composite dark sectors are an obvious target for such investigations, thanks to guidance from both QCD phenomenology as well as lattice QCD calculations with unphysical quark masses, approaching pure-gauge SU(3) Yang--Mills theory as the fermion mass becomes infinite.
See Refs.~\cite{Spergel:1999mh, Faraggi:2000pv, Cline:2013zca, Boddy:2014qxa, Appelquist:2015yfa, Appelquist:2015zfa, Soni:2016gzf, Dienes:2016vei, Forestell:2016qhc, Acharya:2017szw, Berlin:2018tvf, Hochberg:2018rjs, LatticeStrongDynamics:2020jwi, Huang:2020mso, Kang:2021epo, Assi:2023cfo, Batz:2023zef} for representative studies over many years, and Refs.~\cite{Kribs:2016cew, DeGrand:2019vbx} for brief reviews.
While it is well known that the QCD transition is a crossover for physical quark masses, it becomes first order if the physical quarks are replaced by sufficiently light or sufficiently heavy fermions.
In particular, the confinement transition in the SU(3) pure-gauge theory corresponding to infinitely massive fermions is known to be weakly first order, becoming strongly first order for SU($N$) with $N \geq 4$.
Larger values of $N$ appear in many composite dark matter models, including the `dark baryon' of Stealth Dark Matter with even $N \geq 4$~\cite{Appelquist:2015yfa, Appelquist:2015zfa, LatticeStrongDynamics:2020jwi}, and a variety of `dark glueball' models~\cite{Soni:2016gzf, Forestell:2016qhc, Acharya:2017szw, Huang:2020mso, Kang:2021epo, Batz:2023zef}.

Most lattice field theory analyses of these systems~\cite{Kribs:2016cew, DeGrand:2019vbx} employ standard Markov-chain Monte Carlo importance sampling techniques.
However, this approach typically struggles at first-order phase transitions, where it is challenging for Markov-chain updating algorithms to tunnel between the two coexisting phases with different energy densities. 
The exponential suppression of the tunnelling rate as the lattice volume increases towards the thermodynamic limit generically implies super-critical slowing down --- exponentially worsening autocorrelations and increasing computational costs required to obtain a representative sample of field configurations.
This phenomenon is also more severe for stronger phase transitions with larger latent heat.
In recent years a few potential alternative approaches have been explored to avoid this challenge, including parallel tempering~\cite{Borsanyi:2022xml, Borsanyi:2022fub} and density of states methods like the Functional Fit Approach~\cite{Giuliani:2016tlu, Giuliani:2017fss, Gattringer:2020mbf} and the Logarithmic Linear Relaxation (LLR) algorithm~\cite{Langfeld:2012ah, Langfeld:2015fua, Langfeld:2016kty, Korner:2020vjw, Lucini:2023irm}.

In this work we employ the LLR approach, in which the density of states is determined by calculating a piecewise-linear approximation to the slope of its logarithm.
This enables exponential error suppression~\cite{Langfeld:2012ah, Langfeld:2015fua, Langfeld:2016kty}, which is crucial since we need to resolve the density of states across many orders of magnitude in order to study phase transitions.
The LLR algorithm has recently been applied to investigate a variety of lattice systems, including the Ising model, the hexagonal Hubbard model, and gauge theories with gauge groups U(1), SU(2), SU(3) and Sp(4)~\cite{Langfeld:2012ah, Langfeld:2015fua, Langfeld:2016kty, Korner:2020vjw, Langfeld:2022uda, Mason:2022trc, Mason:2022aka, Lucini:2023irm, Mason:2023ixv}.

We apply the LLR algorithm to analyze the bulk phase transition of pure-gauge lattice Yang--Mills theories, considering SU($N$) gauge groups with $4 \leq N \leq 8$.
Preliminary results from this work were presented in Refs.~\cite{Springer:2021liy, Springer:2022qos, Springer:2023wok}.
We begin in the next section by briefly reviewing the phase structure of lattice Yang--Mills theories, contrasting the bulk transition we focus on here against the physical deconfinement transition that we will study in future work.
Section~\ref{sec-LLR} provides a similar review of the LLR approach, which we build on in \secref{sec-sample} by discussing specific algorithmic considerations for large $N \geq 4$.
In \secref{sec-BULK} we present our results from LLR analyses of the bulk transition of SU($N$) Yang--Mills.
Finally we conclude in \secref{sec-conclusion} with a discussion of our planned next steps, including ongoing studies of the SU($N$) deconfinement phase transition.

\section{Transitions of SU($N$) lattice Yang--Mills}
\label{twotrans}
For all SU($N$) gauge groups we use the lattice action
\begin{equation}
  \label{eq:action}
  S = -\frac{\beta}{N} \sum_{x,\mu<\nu} \mathrm{Re}\mathrm{Tr}\left[U_{\mu\nu}(x)\right],
\end{equation}
based on the plaquette $U_{\mu\nu}(x) = U_{\mu}(x) U_{\nu}(x+\hat{\mu}) U_{\mu}^{\dag}(x+\hat{\nu}) U_{\nu}^{\dag}(x)$.
Here the sum runs over all lattice sites $x$, and $U_{\mu}(x)$ is the SU($N$)-valued link variable attached to site $x$ in direction $\hat{\mu}$.
Smaller values of the lattice parameter $\beta$ correspond to stronger bare Yang--Mills couplings $g_0^2$ at the cut-off scale set by the lattice spacing.
For any finite system, this action differs from the standard Wilson action by an irrelevant constant. 
We employ it to simplify the relation between $S$ and the average plaquette
\begin{equation}
  u_P = \frac{1}{6NV} \sum_{x,\mu<\nu} \mathrm{Re}\mathrm{Tr}\left[U_{\mu\nu}(x)\right] \in [0, 1],
\end{equation}
namely $S = -6V\beta u_P$, with $V$ the volume of the space-time lattice.
Further we define the energy of a given lattice configuration as $E=\frac{S}{\beta}=-6V u_P$, which we will see again in the next section.

For SU($N$) Yang--Mills theories on an $N_s^3 \times N_t$ lattice, there are two distinct transitions we could consider.
The physically relevant deconfinement transition corresponds to the spontaneous breaking of the $Z_N$ center symmetry, with the Polyakov loop the corresponding order parameter.
This transition is first order for $N \geq 3$, but only weakly so for $N = 3$, in the sense that the latent heat $L_h$ for the SU(3) deconfinement transition is significantly smaller than would be expected from the $L_h \propto N^2$ scaling observed for larger $N \geq 4$~\cite{Lucini:2012gg}.
For any given $N_t$, the critical temperature $T_c = 1 / (\alat N_t)$ corresponds to a critical $\beta_c(N_t)$ that determines the lattice spacing $\alat$.
Since we can identify this $T_c$ with the physical deconfinement temperature, we can see that $\beta_c \to \infty$ as $N_t \to \infty$ in the $\alat \to 0$ continuum limit.
Determining the latent heat $L_h$ from the jump in the average plaquette at the lattice transition requires also evaluating how the lattice parameter $\beta$ depends on the lattice spacing~\cite{Lucini:2005vg}: 
\begin{equation}
  \label{eq:latentheat}
  \frac{L_h}{T_c} = N_t^4 \alat \frac{\partial \beta}{\partial \alat} 6 \Delta u_P.
\end{equation}

In addition, there can be a much stronger `bulk' phase transition that occurs at an $N_t$-independent coupling $\beta_{\text{bulk}}$.
That is, this bulk transition is not a feature of the $\alat \to 0$ continuum limit.
It can be described in terms of the condensation of $Z_N$ monopoles~\cite{Brower:1981rz}, but is easiest to see directly in the average plaquette $u_P$.
With our action \eq{eq:action}, $N \geq 5$ is needed to obtain a first-order transition (weakly first-order for $N = 5$), with a continuous crossover for smaller $2 \leq N \leq 4$~\cite{Lucini:2005vg}.
However, it has long been known~\cite{Bhanot:1981eb} that first-order bulk transitions appear for all $N \geq 2$ in the extended $\{\beta, \beta_A\}$ parameter space of the fundamental--adjoint action
\begin{equation}
  \begin{split}
    S_A & = -\frac{\beta}{N} \sum_{x,\mu<\nu} \mathrm{Re}\mathrm{Tr}\left[U_{\mu\nu}(x)\right] - \frac{\beta_A}{N} \sum_{x,\mu<\nu} \mathrm{Re}\mathrm{Tr_A}\left[U_{\mu\nu}(x)\right] \\
        & = -\frac{\beta}{N} \left(\sum_{x,\mu<\nu} \mathrm{Re}\mathrm{Tr}\left[U_{\mu\nu}(x)\right] + r \sum_{x,\mu<\nu} \mathrm{Re}\mathrm{Tr_A}\left[U_{\mu\nu}(x)\right] \right), \label{eq:fundadj}
  \end{split}
\end{equation}
where $\mathrm{Tr_A}$ is the trace in the adjoint representation and we define the ratio $r = \frac{\beta_A}{\beta}$.
Our \eq{eq:action} corresponds to the $\beta_A = 0$ line in this extended parameter space.
For $N = 2$, the transition is only first-order for relatively large values of $\beta_A \gtrsim 1.25$~\cite{Lucini:2013wsa}, which decrease as $N$ increases, becoming negative for $N \geq 5$.
Even when the `transition' is really a continuous crossover, it leads to large lattice artifacts~\cite{Hasenbusch:2004yq, Hasenfratz:2011xn}, motivating the development of bulk-preventing actions, for example \refcite{Rindlisbacher:2023hhn}.

When both transitions are first order, for example with $N \geq 5$ using the $\beta_A = 0$ action \eq{eq:action} that we consider in this work, the bulk transition features a much larger latent heat compared to the deconfinement transition~\cite{Lucini:2005vg}.
This makes the bulk transition a useful target for algorithmic testing and development of the sort we present here, which will provide a foundation for subsequent application to the deconfinement transition that persists in the physical continuum limit.
In particular, these strong bulk transitions with large latent heat are precisely the domain in which traditional Markov-chain methods encounter the difficulties described in \secref{intro}, making them an excellent proving ground for density of states approaches and the LLR algorithm in particular, to which we now turn.

\section{Brief review of Logarithmic Linear Relaxation}
\label{sec-LLR}
Generic observables in lattice field theory are defined through the euclidean path integral,
\begin{align}
  \label{eq:obs}
  \vev{\cO} & = \frac{1}{Z} \int \cD\phi \, \cO(\phi) \, e^{S[\phi]} &
  Z & = \int \cD \phi \, e^{S[\phi]},
\end{align}
where $S[\phi]$ is the lattice action --- \eq{eq:action} in our case.
Explicitly solving path integrals is only possible in very special cases.
Standard Monte Carlo techniques instead sample only a small number of representative field configurations, with probability $\propto e^{S[\phi]}$, to obtain systematically improvable approximate results.

If we had access to the density of states
\begin{equation}
  \rho(E) = \int \cD \phi \, \de(S[\phi] - \beta E),
\end{equation}
then for observables that depend only on the action \eq{eq:obs} would simplify to a one-dimensional integral over the energy:
\begin{align}
  \vev{\cO(\beta)} & = \frac{1}{Z(\beta)} \int \d E \, \cO(E) \, \rho(E) \, e^{\beta E} &
  Z(\beta) & = \int \d E \, \rho(E) \, e^{\beta E}.
\end{align}
See \refcite{Langfeld:2015fua} for discussions of more general observables.
In practice, the density of states $\rho(E)$ varies over hundreds or thousands of orders of magnitude and is difficult to determine with sufficient precision in a straightforward manner.
The LLR algorithm provides a solution to this problem~\cite{Langfeld:2012ah, Langfeld:2015fua}.

The first step in the LLR approach is to divide the energy range of interest into a number of small energy intervals of size $\de$.
These energy intervals need to be small enough for the logarithm of the density of states to be piecewise linear in the energy: $\log \rho(E) \approx aE$, or equivalently $\rho(E) \approx e^{aE}$, where the new parameter `$a$' is not to be confused with the lattice spacing $\alat$.
We next introduce a restricted expectation value that samples only field configurations with energies in a given interval centered at the fixed energy value $E_i$:
\begin{align}
  \vev{\vev{\cO}}_{E_i,\de}(a) & = \frac{1}{\cN}\int \cD \phi \, \cO(E) \, \theta_{E_i,\de} \, e^{-\frac{a}{\beta}S[\phi]} = \frac{1}{\cN} \int_{\Emin}^{\Emax} \d E \, \cO(E) \, \rho(E) \, e^{-aE}, \label{Heaviside} \\
  \cN & = \int \cD \phi \, \theta_{E_i,\de} \, e^{-\frac{a}{\beta}S[\phi]} = \int_{\Emin}^{\Emax} \d E \, \rho(E) \, e^{-a E}. \label{norm}
\end{align}
Here the modified Heaviside function $\theta_{E_i,\de}$ vanishes for all energies outside of the interval from $\Emin = E_i - \de / 2$ to $\Emax = E_i + \de / 2$.
In this restricted expectation value, we have moved the LLR parameter $a$ into the Boltzmann weight (with an irrelevant negative sign).

Our goal is to determine $a(E_i,\de) = \left.\frac{\d \log \rho(E)}{\d E}\right|_{E_i,\de}$ using a particular restricted expectation value in \eq{Heaviside}, which will allow us to numerically reconstruct $\rho(E)$ itself with exponential error suppression~\cite{Langfeld:2012ah, Langfeld:2015fua, Langfeld:2016kty}.
We can do this by considering $\cO(E) = (E - E_i)$ and noting that $\vev{\vev{E - E_i}}_{\de} \approx 0$ when we find the value of $a$ that makes $\rho(E) e^{-a E}$ an approximately uniform distribution within the small energy interval.
As a consistency check, we can expand the restricted expectation value in powers of the small interval size \de to confirm that we recover our initial `logarithmic linear' assumption $\log \rho(E) \approx aE$:
\begin{align}
  \vev{\vev{E - E_i}}_{\de}(a) & = \frac{1}{\cN} \int_{\Emin}^{\Emax} \d E \, (E-E_i) \, \rho(E) \, e^{-aE} \\
   = \frac{1}{\cN} \frac{\de}{2} & \left[\left(\frac{\de}{2}\right) e^{-a\left(E_i+\frac{\delta}{2}\right)} \rho\left(E_i+\frac{\de}{2}\right) + \left(-\frac{\de}{2}\right) e^{-a\left(E_i-\frac{\de}{2}\right)} \rho\left(E_i-\frac{\de}{2}\right)\right] + \cO(\de^3) = 0. \nonumber
\end{align}
Further expanding $e^{\pm a \frac{\de}{2}}$ and $\rho(E_i\pm \frac{\de}{2})$, then taking $\de \to 0$, gives us the desired result:
\begin{align}
  0 & = \left(\rho(E_i) + \frac{\de}{2} \frac{\d \rho(E)}{\d E}\Bigr|_{E=E_i}\right)\left(1-a\frac{\de}{2}\right) - \left(\rho(E_i)-\frac{\de}{2}\frac{\d \rho(E)}{\d E}\Bigr|_{E=E_i}\right)\left(1+a\frac{\de}{2}\right) \nonumber \\
  & = \left(-\rho(E_i)a + \frac{\d \rho(E)}{\d E}\Bigr|_{E = E_i} - \rho(E_i)a + \frac{\d \rho(E)}{\d E}\Bigr|_{E=E_i}\right)\frac{\de}{2} \\
  \implies a & = \frac{1}{\rho(E_i)}\frac{\d \rho(E)}{\d E}\Bigr|_{E=E_i} = \frac{\d \log \rho(E)}{\d E}\Bigr|_{E=E_i}.
\end{align}
To reiterate: The LLR parameter $a(E_i)$ is a linear approximation of the energy derivative of the logarithm of the density of states $\rho(E)$, evaluated at $E_i$.
By numerically integrating over $a(E)$ for all relevant energy intervals $E_i$ we can determine the normalizable probability density $P_{\beta}(E) = \rho(E) e^{\beta E}$ with exponential error suppression~\cite{Langfeld:2012ah, Langfeld:2015fua, Langfeld:2016kty}.
We will say more about this step in \secref{sec-BULK}.
The two-peak structure in $P_{\beta}(E)$ that characterizes a first-order transition corresponds to non-monotonic energy dependence in $a(E)$, which is nicely illustrated by Fig.~9 in \refcite{Lucini:2023irm}.

Solving $\vev{\vev{E - E_i}}_{\de}(a) = 0$ for $a$ is a tractable problem~\cite{Langfeld:2015fua} that can be tackled with standard stochastic root-finding algorithms such as the iterative Newton--Raphson (NR) method:
\begin{equation}
  a_{j+1} = a_j + \frac{\vev{\vev{E - E_i}}_{\de}(a_j)}{\si_{\de}^2(a_j)}.
  \label{eq:newraph}
\end{equation}
Starting from some initial $a_0$, each iteration that updates $a_j$ to $a_{j+1}$ requires evaluating both the restricted expectation value $\vev{\vev{E - E_i}}_{\de}(a_j)$ and the restricted variance $\si_{\de}^2(a_j)$.
This is done using restricted importance sampling, which we discuss in more detail in the next section.

Two aspects of this restricted importance sampling are worth mentioning here, before the more detailed discussion to come.
First, although $\vev{\vev{E - E_i}}_{\de}(a_j) = 0$ corresponds to a fixed point of \eq{eq:newraph}, stochastic fluctuations in the Monte Carlo evaluation of the restricted expectation value means that the iterative process will fluctuate around this fixed point rather than exactly converging to it.
These fluctuations can be mitigated by introducing under-relaxation~\cite{Langfeld:2015fua},
\begin{equation}
  a_{j+1} = a_j + \frac{1}{j+1}\frac{\vev{\vev{E - E_i}}_{\de}(a_j)}{\si_{\de}^2(a_j)},
  \label{eq:robmon}
\end{equation}
corresponding to the Robbins--Monro (RM) algorithm.
Depending on how much $a_j$ needs to evolve from its initial value, this under-relaxation may be too aggressive, which we address by first carrying out 30 iterations using \eq{eq:newraph} and only then turning on under-relaxation. 
In addition, we terminate each stochastic root-finding procedure after only 30+20 NR+RM iterations, running $\Nj = 5$ independent calculations for each energy interval and performing jackknife uncertainty analyses using the final values of $a_{50}$.

Second, if $a_j$ is sufficiently close to the correct value, we can expect approximately uniformly distributed sampling of $(E - E_i)$ within the small energy interval.
This motivated us to experiment with jackknife analyses using the \Nj values of $a_j$ that produce the most-uniform distribution of $(E - E_i)$ measurements, as quantified by either the Kolmogorov--Smirnov test or the Hellinger fidelity.
However, this gave us no improvement over simply using the final $a_{50}$, most likely due to fluctuations in the statistical sampling.
Another step we can take is to approximate the variance as $\si_{\de}^2(a_j) \approx \frac{\de^2}{12}$ for the uniform distribution, which is done in most prior work~\cite{Langfeld:2012ah, Langfeld:2015fua, Korner:2020vjw, Langfeld:2022uda, Lucini:2023irm, Mason:2023ixv}.
However, depending on the value of $a_j$ and the approach used to restrict the importance sampling to the small energy interval, this approximation may be a poor one, which we have observed to cause difficulties in the stochastic root finding (potentially relevant to \refcite{Korner:2020vjw}). 
At the same time, we have also observed that $\si_{\de}^2(a_j)$ can be underestimated on occasion when evaluated with only a limited number of Monte Carlo samples, leading to unreasonably large fluctuations in $a_{j+1}$.
Our preferred approach to resolve these difficulties is to use $\max\left\{\si_{\de}^2(a_j), \frac{\de^2}{12}\right\}$ in the denominator of Eqs.~\ref{eq:newraph} and \ref{eq:robmon}.

\section{Algorithmic considerations for large $N$}
\label{sec-sample}
The previous two paragraphs have already commented on some generic aspects of the stochastic root finding employed within the LLR algorithm.
In this section we discuss in more detail our experiments with algorithms to carry out the restricted importance sampling used to compute $\vev{\vev{E - E_i}}_{\de}(a_j)$ and $\si_{\de}^2(a_j)$ in Eqs.~\ref{eq:newraph} and \ref{eq:robmon}, in the specific context of large-$N$ lattice calculations.
The task is to carry out importance sampling based on the probability weight $e^{-a_j S}$ while constraining the energy to remain within --- or at least near --- the interval $E_i \pm \de/2$.
(Without loss of generality, we set $\beta = 1$ in Eqs.~\ref{Heaviside}--\ref{norm}.)
Most previous work~\cite{Langfeld:2012ah, Langfeld:2015fua, Lucini:2023irm, Mason:2023ixv} employs traditional over-relaxed Cabibbo--Marinari `quasi-heatbath' (QHB) updates on SU(2) sub-groups~\cite{Cabibbo:1982zn}, imposing hard cut-offs on the energy either as part of the update or part of the accept/reject test.
Since SU(2) over-relaxation updates are micro-canonical, some parallelization can be introduced via domain decomposition, despite the constraint on the global energy~\cite{Lucini:2023irm}.
In addition, calculations for different energy intervals are completely independent and can be run in parallel.

For large $N \geq 4$, \refcite{deForcrand:2005xr} argues that performing over-relaxation updates on the full SU($N$) gauge links is significantly more efficient than the QHB approach, in terms of the computational cost required to decorrelate Polyakov loop measurements.
Such full-SU($N$) over-relaxation updates are no longer micro-canonical, so they don't have to be combined with heatbath updates.
Although originally introduced long ago by Creutz~\cite{Creutz:1987xi}, this method historically suffered from low acceptance rates until being improved by Refs.~\cite{Kiskis:2003rd, deForcrand:2005xr}.

We have implemented full-SU($N$) over-relaxation updates as an alternative to the QHB approach.
Unfortunately, imposing hard cut-offs on the global energy prevents parallelization within each energy interval and re-introduces difficulties with low acceptance rates, especially as the energy interval size \de decreases.
To address both of these issues, we have also implemented hybrid Monte Carlo (HMC) updates, in which new field configurations are generated by carrying out approximate molecular dynamics (MD) updates along a trajectory of length $\Delta \tau = 1$ in a fictitious `MD time' $\tau$.
HMC updates are easily parallelizable, and offer control over acceptance rates by adjusting the number of steps $N_{\text{step}} = \Delta \tau / \de_{\tau}$ into which the MD trajectory is divided.
Although unconstrained local updating algorithms exhibit much more computationally efficient decorrelation compared to HMC~\cite{Gupta:1988yw}, the need to impose global constraints on the energy in the LLR algorithm makes the HMC algorithm more competitive.

\begin{figure}[tbp]
  \includegraphics[width=0.475\linewidth]{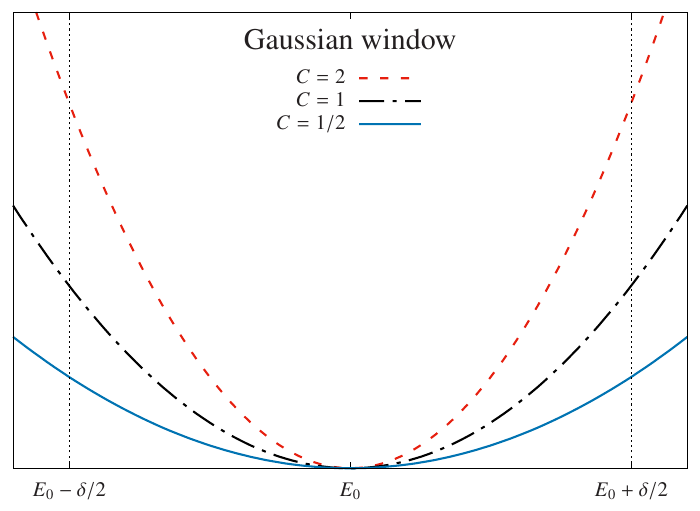}\hfill
  \includegraphics[width=0.475\linewidth]{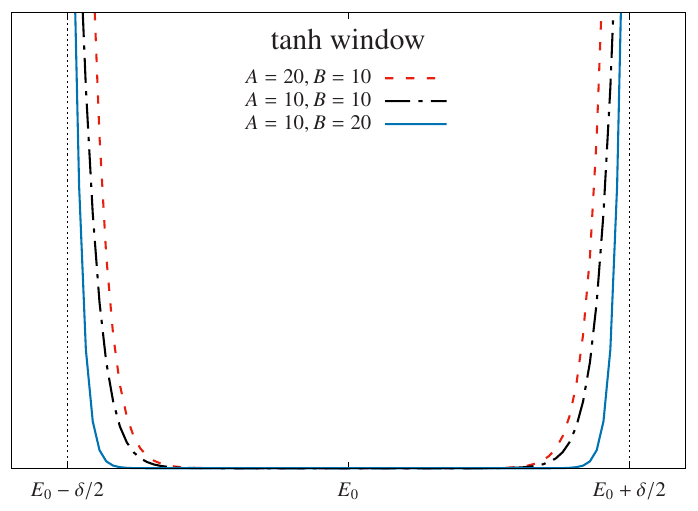}
  \caption{Illustrations of differentiable window functions that can be used in the HMC algorithm to keep the energy near the interval $E_0 \pm \de/2$, involving a simple Gaussian (left) or a difference of $\tanh$ functions (right).  While the $\tanh$ windows better approximate the hard cut-off used in most previous work, they produce large forces that lead us to use the Gaussian window with $C = 1$.}
  \label{fig:windows}
\end{figure}

The complication is that in order to control HMC acceptance rates, the energy constraint needs to be incorporated into the MD evolution, and this requires replacing hard cut-offs by a differentiable `window function'.
The standard choice~\cite{Langfeld:2016kty, Korner:2020vjw} is to introduce a Gaussian window function,
\begin{align}
  \label{eq:gauss}
  \vev{\vev{E - E_i}}_{\de}(a) & = \frac{1}{\cN} \int \d E \, (E-E_i) \, e^{-W(E, E_i, \de)} \, \rho(E) \, e^{-aE}, \\
                               & W(E, E_i, \de) = C\frac{(E - E_i)^2}{2\de^2}, \nonumber
\end{align}
and similarly for $\cN$, where $C$ is a tunable parameter that allows the width of the Gaussian to differ from the interval size $\de$.
Another possibility that we have explored is to use a $\tanh$ window function with two tunable parameters $A$ and $B$,
\begin{equation}
  \label{eq:tanh}
  W(E, E_i, \de) = A\left(1 + \frac{1}{2}\left[\tanh(B(E - \Emax)) - \tanh(B(E - \Emin))\right]\right).
\end{equation}
In this approach, instead of explicitly restricting the integration to $\Emin \leq E \leq \Emax$ as in Eqs.~\ref{Heaviside}--\ref{norm}, excursions away from this interval are allowed but suppressed by the effective probability weight $e^{-(aE + W)}$.

In \fig{fig:windows} we illustrate these two window functions.
While the $\tanh$ windows better approximate the hard cut-off used in most previous work, they produce large forces around the edges of the interval, $E_i\pm \frac{\de}{2}$, which hurts the performance of the algorithm.
Larger values of $C$ similarly increase the forces, leading us to use the Gaussian window function with $C = 1$ in our calculations.
The challenge we encounter with smaller $C$ is that larger excursions away from the small energy interval are possible.
As mentioned at the end of \secref{sec-LLR}, this can lead to sampling of $(E - E_i)$ that is not approximately uniform within the small energy interval, even for the correct value of $a$ that successfully produces $\vev{\vev{E - E_i}}_{\de}(a) = 0$.
This was our motivation for introducing $\max\left\{\si_{\de}^2(a_j), \frac{\de^2}{12}\right\}$ in the NR and RM iterations (Eqs.~\ref{eq:newraph} and \ref{eq:robmon}), which can significantly improve the performance of these root-finding methods when we employ HMC importance sampling to compute $\vev{\vev{E - E_i}}_{\de}(a_j)$ and $\si_{\de}^2(a_j)$. 

So far we have described three restricted importance sampling algorithms that can be used to find the LLR parameter $a = \frac{\d \log \rho(E)}{\d E}$ through a combination of NR and RM iterations, Eqs.~\ref{eq:newraph} and \ref{eq:robmon}.
These are over-relaxed QHB updates on SU(2) subgroups, full-SU($N$) over-relaxation updates, and the HMC algorithm, with hard energy cut-offs in the first two cases and a differentiable window function in the third.
We have implemented the LLR algorithm using all three options and confirmed that they produce consistent results for $a(E)$.
As a further check, we have also implemented and tested a fourth option of naive Metropolis--Rosenbluth--Rosenbluth--Teller--Teller (MRRTT) updates generalized from the SU(3) case considered by \refcite{Katznelson:1984kw}.
The HMC-based algorithm that we use for our main study presented in \secref{sec-BULK} is available in public code based on the MILC software~\cite{LargeN_YM_code}, while we implemented other algorithms in our fork of S.~Piemonte's \texttt{LeonardYM} package~\cite{LeonardYM_code}.

\begin{figure}[btp]
  \includegraphics[width=0.475\linewidth]{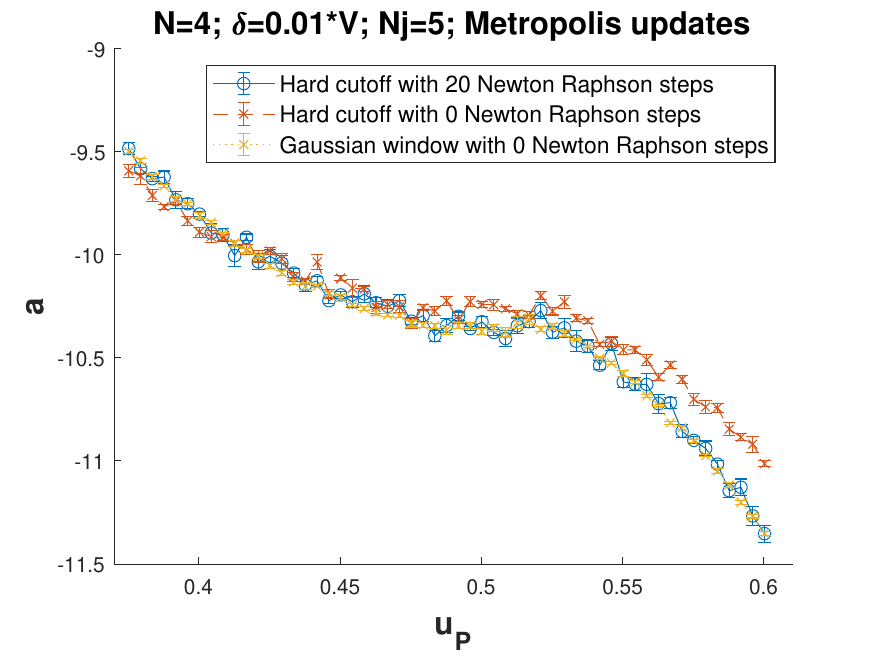}\hfill
  \includegraphics[width=0.475\linewidth]{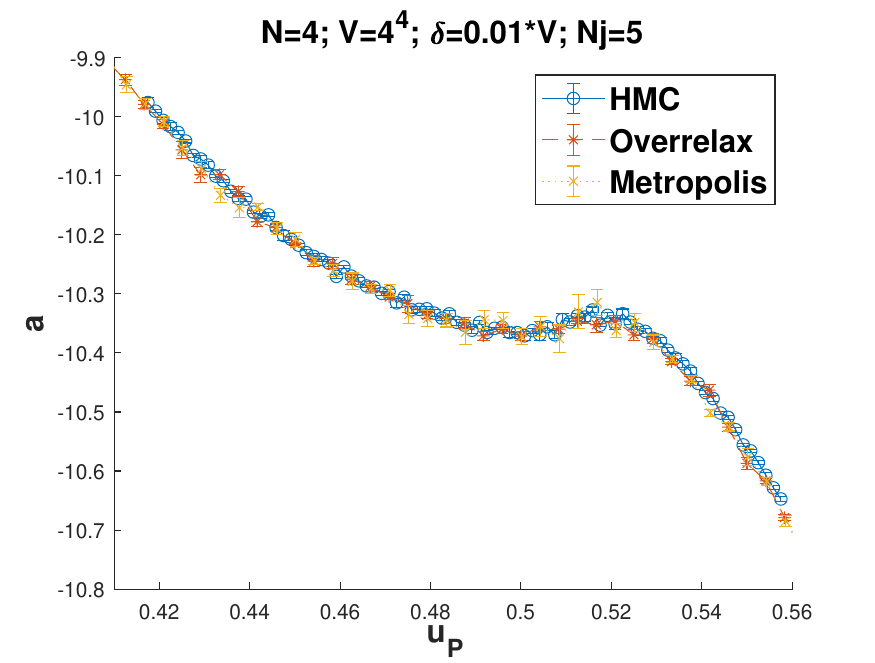}
  \caption{SU(4) results for the LLR parameter $a$ from $V=4^4$ lattices with an energy interval size of $\de = 0.01V$, vs.\ the average plaquette. Statistical uncertainties are obtained by performing $\Nj = 5$ independent calculations per interval. \textbf{Left:} Comparing hard energy cut-offs (with and without NR iterations before turning on under-relaxation) and the Gaussian window function, in all cases using naive MRRTT updates.  \textbf{Right:} Comparing HMC, full-SU($N$) over-relaxation, and MRRTT updates, in all cases using the Gaussian window function without any initial NR iterations.  The orange points are the same in both plots, with different axis ranges.}
  \label{fig:compareupdate}
\end{figure}

Figure~\ref{fig:compareupdate} demonstrates the consistent results for $a(u_P)$ that we obtain for these different algorithms, at least when we take care not to under-relax too aggressively.
Recall that the average plaquette is $u_P = -E / 6V \in [0, 1]$.
For these small tests we consider SU(4) Yang--Mills theory on $4^4$ lattices with an energy interval size of $\de = 0.01V$.
In the left panel we consider only naive MRRTT updates, using either hard energy cut-offs or the Gaussian window function.
The results are mostly in agreement within their statistical uncertainties from jackknifing $\Nj = 5$ independent calculations in each small energy interval.
Discrepancies for relatively large $u_P \gtrsim 0.55$ are resolved by beginning the stochastic root finding with 20 NR iterations in the hard cut-off case, instead of under-relaxing from the start.
In the right panel of \fig{fig:compareupdate} we compare results obtained using HMC, full-SU($N$) over-relaxation, and MRRTT updates, in all three cases using the Gaussian window function without any initial NR iterations, and finding full agreement.
In addition to these SU(4) $4^4$ tests, we have also carried out a smaller number of checks on larger volumes up to $8^4$ and with larger $N = 6$, in all cases finding consistency among the various algorithmic options.

\begin{figure}[btp]
  \includegraphics[width=0.475\linewidth]{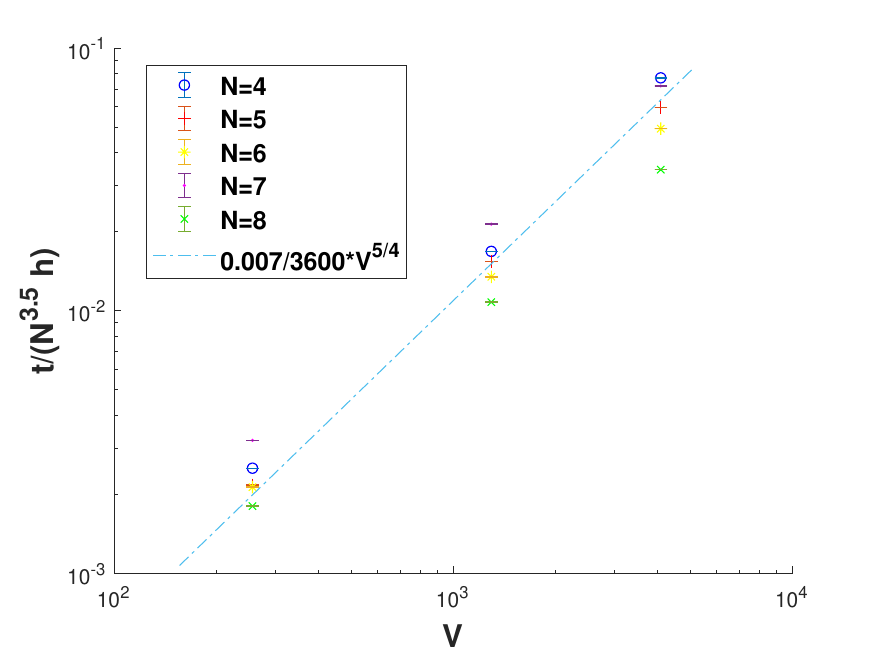}
  \caption{Computational costs in core-hours for a single energy interval of size $\de=0.01V$, involving 30+20 NR+RM iterations, each with 60 HMC trajectories.  We compare all $4 \leq N \leq 8$ with the y-axis normalized by the expected $N^{7/2}$ dependence, and consider the three volumes $V=4^4 = 256$, $6^4 = 1296$ and $8^4 = 4096$, with the dot-dashed line displaying the expected volume dependence $\propto V^{5/4}$ on log--log axes.  Although the costs are dominated by the 3000 HMC trajectories, they also include initialization of a gauge configuration in the energy interval as discussed in the text.}
  \label{fig:runtime}
\end{figure}

\begin{table}[btp]
  \centering
  \renewcommand\arraystretch{1.25} 
  \addtolength{\tabcolsep}{3 pt}   
  \begin{tabular}{ |c|c|c|c|c|c|c| }
    \hline
          & \multicolumn{2}{|c|}{$4^4$} & \multicolumn{2}{|c|}{$6^4$} & \multicolumn{2}{|c|}{$8^4$} \\\hline
    SU(4) & $0.32201(36)$ & 96\%        & $2.1569(37)$   & 96\%       & $9.887(39)$  & 95\%         \\\hline
    SU(5) & $0.6095(13)$  & 95\%        & $4.297(11)$    & 95\%       & $16.670(38)$ & 92\%         \\\hline
    SU(6) & $1.128(50)$   & 94\%        & $7.144(0.018)$ & 93\%       & $26.176(60)$ & 88\%         \\\hline
    SU(7) & $1.7007(23)$  & 93\%        & $11.341(35)$   & 92\%       & $38.05(10)$  & 87\%         \\\hline
    SU(8) & $2.6133(25)$  & 90\%        & $15.631(21)$   & 90\%       & $50.00(13)$  & 88\%         \\\hline
  \end{tabular}
  \caption{\label{tab_runtime}The computational costs in core-hours shown in \fig{fig:runtime}, along with the corresponding HMC acceptance rates, for a single energy interval with different volumes and SU($N$) gauge groups.}
\end{table}

The HMC updates produce the most precise results in \fig{fig:compareupdate}.
This, along with their straightforward data-parallelism and control over acceptance rates, led us to use the HMC algorithm for all results shown in the remainder of this paper.
Before turning to those results, in \fig{fig:runtime} we confirm that computational costs for our overall LLR calculations scale as we would expect for the HMC algorithm.
These expectations are that costs need to scale $\propto V^{5/4}$ and $\propto N^{7/2}$ in order to keep acceptance rates fixed.
In \fig{fig:runtime} we normalize the $4 \leq N \leq 8$ data by $N^{7/2}$ and compare them with the dot-dashed line $\propto V^{5/4}$ on log--log axes, observing consistency with both expected scalings.
Small deviations are to be expected due to both variations in the acceptance rates (shown in Table~\ref{tab_runtime}) and our procedure to initialize a gauge configuration in the energy interval of interest.
Starting from some initial configuration and $\beta = 1$, we run (unconstrained) over-relaxed QHB updates and after every fifth sweep we increase or decrease $\beta$ depending on whether the energy is too high or too low, respectively.
This procedure terminates once the energy is within the target interval, or $\beta$ has been adjusted 2000 times, which unpredictably affects the computational costs shown in \fig{fig:runtime} and Table~\ref{tab_runtime}.
We also use the final value of $\beta$ as an initial guess for the LLR parameter $-a_0$, resetting $\beta = 1$.
We have confirmed that our procedure produces consistent results independent of the initial configuration, in particular comparing the options of setting all links to unit matrices or to random unitary matrices.

\section{Bulk transition results}
\label{sec-BULK}
In this section we present our results for the bulk phase transition, confirming a first-order transition for $N \geq 5$ and determining the corresponding $\beta_{\text{bulk}}$ and plaquette jump $\Delta u_P$.
Since many algorithmic possibilities were reviewed above, we begin by summarizing the LLR setup we used to obtain the following results.
The most important choice we have made, for reasons discussed above, is to use the HMC algorithm with the $C = 1$ Gaussian window function (\eq{eq:gauss}) to compute the restricted expectation value $\vev{\vev{E - E_i}}_{\de}(a_j)$.
We carried out these calculations using just $\de^2$ in the NR and RM iterations (Eqs.~\ref{eq:newraph} and \ref{eq:robmon}), and only identified the benefits of using $\max\left\{\si_{\de}^2(a_j), \frac{\de^2}{12}\right\}$ in the course of this work. 
Instead, to control occasional instabilities we impose a maximum on the amount the LLR parameter can change in each iteration, $|a_{j+1} - a_j| \leq \frac{200}{\de^2}$.
We fix $\de = 0.01V$ and carry out 30 NR iterations followed by 20 RM iterations.
The entire procedure for each energy interval is repeated $\Nj=5$ times, with statistical uncertainties obtained by jackknifing these \Nj independent calculations.
For all of them we carry out the initialization procedure described above, starting with all links set to unit matrices; the \Nj different sequences of pseudo-random numbers lead to different initial guesses $a_0$ at the start of the NR iterations.

\begin{figure}[btp]
  \includegraphics[width=0.475\linewidth]{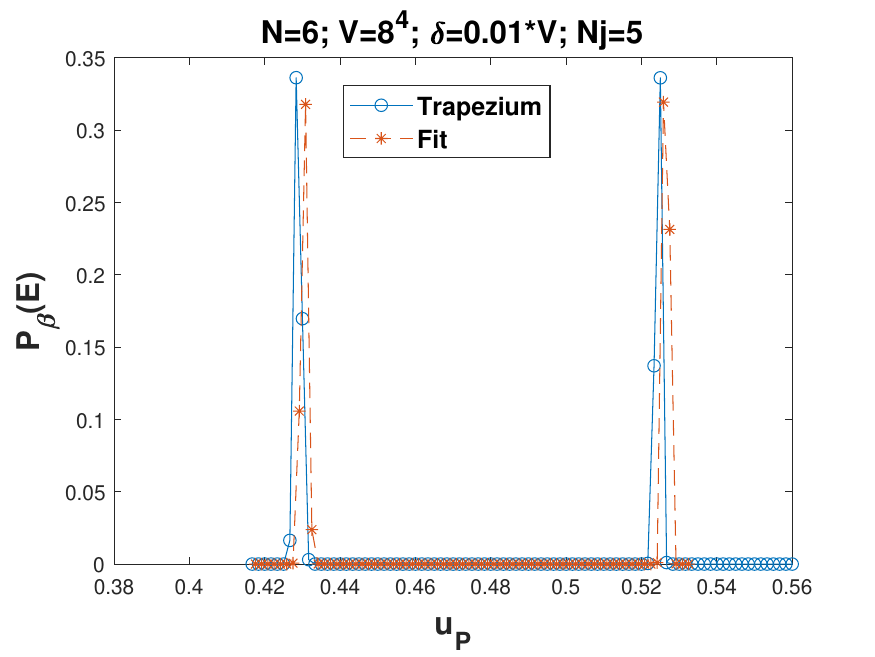}\hfill
  \includegraphics[width=0.475\linewidth]{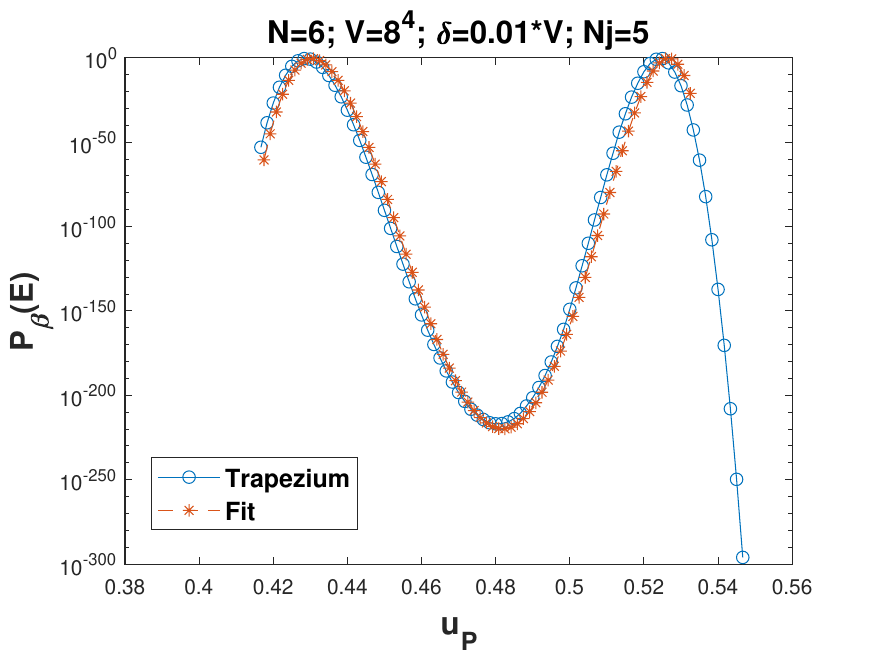}
  \caption{The SU($6$) $8^4$ probability density $P_{\beta}(u_P)$ (omitting uncertainties) reconstructed using both trapezium-rule integration with $\beta = 24.3921952$ and a polynomial fit with $\beta = 24.3914534$, and plotted on either a linear (left) or logarithmic (right) scale.}
  \label{fig:fitvstrap}
\end{figure}

We have compared two ways to reconstruct the normalizable probability density $P_{\beta}(u_P) = \rho(u_P) e^{-6V\beta u_P}$ from our LLR results for $a(u_P)$: Simple trapezium-rule numerical integration and a polynomial fit technique first used in~\cite{Francesconi:2019nph, Francesconi:2019aet}.
For the representative case SU(6) $8^4$, \fig{fig:fitvstrap} compares the results from these two methods, which predict (within statistical uncertainty) the same $\Delta u_P = 0.097(2)$ and $\beta_{\text{bulk}}$ values that differ by only 0.003\% (roughly twice the statistical uncertainty).
The main difference is a slight shift in the $u_P$ values of the two peaks.
Our results below come from using trapezium-rule integration. 
The systematic uncertainty introduced by this choice of reconstruction technique appears to be comparable to the statistical uncertainty.

The two plots in \fig{fig:fitvstrap} show the same results plotted on either a linear or logarithmic scale.
The latter highlights the extremely small probabilities characterizing the valley between the two peaks, which decrease $\propto e^{-\sigma V}$ where $\sigma$ is the surface tension~\cite{Langfeld:2016kty} (an important non-perturbative input for gravitational-wave analyses).
This suppresses the tunneling rate between the two coexisting phases as $V$ and $N$ increase, leading to the difficulties faced by importance sampling Monte Carlo analyses, reviewed in \secref{intro}.
We also encounter the practical complication of results that `underflow' the $10^{-308}$ limit of double-precision numbers.
We use \texttt{MATLAB}'s variable-precision arithmetic (\texttt{vpa}) package to reliably compute $P_{\beta}(u_P) \ll 10^{-1000}$, but limit the range of the y-axis in the figures below to the double-precision domain.

\begin{figure}[btp]
  \includegraphics[width=0.475\linewidth]{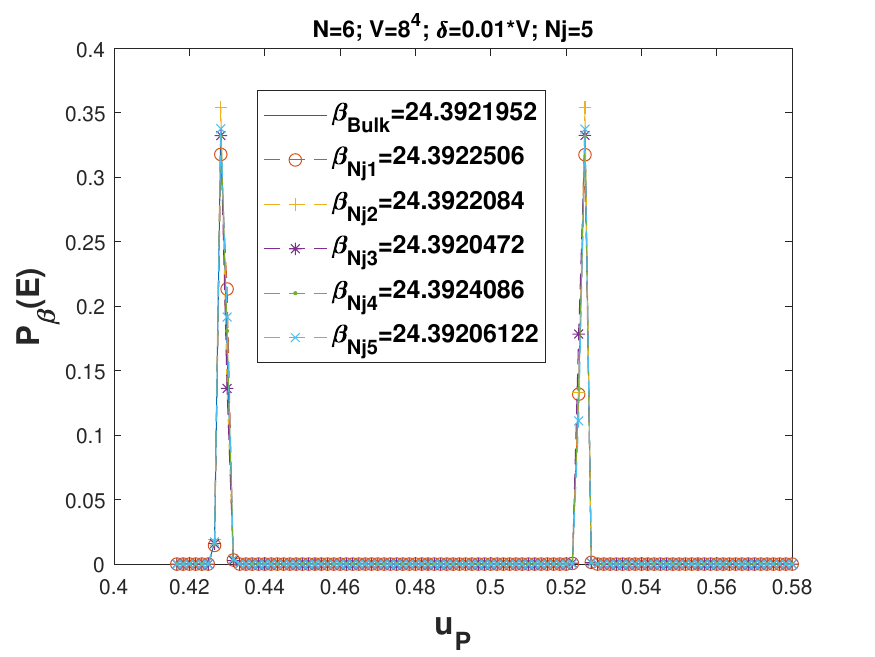}\hfill
  \includegraphics[width=0.475\linewidth]{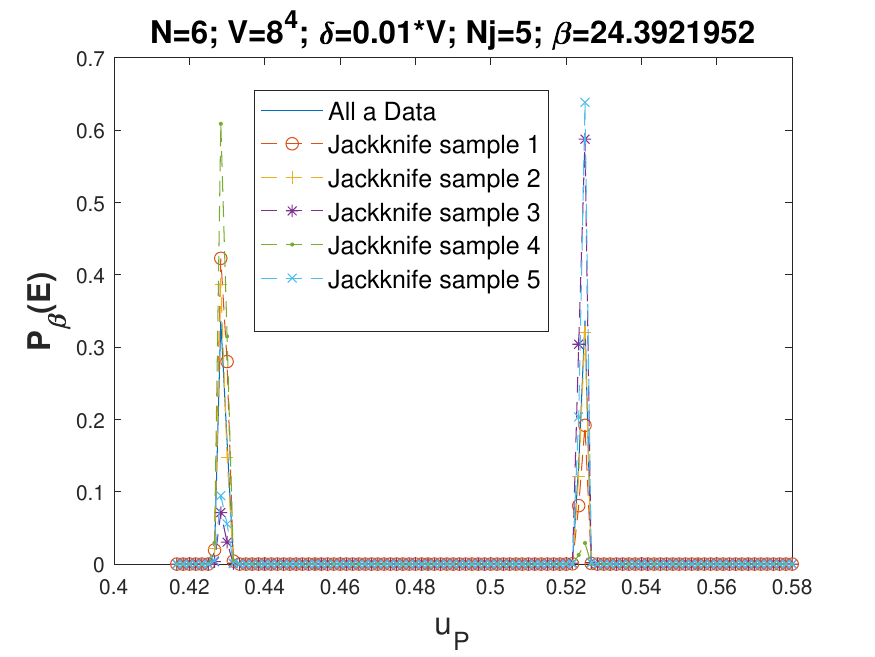}
  \caption{Illustration of our jackknife procedure to estimate statistical uncertainties, considering the SU($6$) $8^4$ case. \textbf{Left:} The probability density $P_{\beta}(u_P)$ calculated for each jackknife sample at the corresponding $\beta_{\text{Nj}i}$ tuned to obtain two equal-height peaks.  The result for the full data set at its $\beta_{\text{bulk}} = 24.3921952$ is also included.  \textbf{Right:} $P_{\beta}(u_P)$ for each jackknife sample, along with that for the full data set, now all evaluated at $\beta_{\text{bulk}}$.}
  \label{fig:errorbeta}
\end{figure}

In \fig{fig:errorbeta} we illustrate the jackknife procedure we use to obtain the statistical uncertainties mentioned above, again using SU(6) $8^4$ as a representative example.
For each jackknife sample of $a(u_P)$ obtained by eliminating one of the \Nj independent calculations in each small energy interval, we adjust $\beta$ so as to obtain two peaks of equal height in the probability density $P_{\beta}(u_P)$.
Because the peaks are both so narrow, they also have equal area to a very good approximation.
The left panel of \fig{fig:errorbeta} shows these five sets of results and the corresponding jackknife estimates for $\beta_{\text{Nj}i}$, which together produce $\beta_{\text{bulk}} = 24.39220(26)$.
The right panel illustrates what would happen if we tried to assign uncertainties to $P_{\beta}(u_P)$ by averaging over the jackknife samples with fixed $\beta$.
These five sets of results differ by up to an order of magnitude, resulting in very large uncertainties.
Since we are interested only in $\beta_{\text{bulk}}$ and $\Delta u_P$, we omit uncertainties on $P_{\beta}(u_P)$ itself in the plots below. 

Returning to the left panel of \fig{fig:errorbeta}, we can read off jackknife samples for the plaquette jump $\Delta u_P$ as the distance between the two peaks after adjusting $\beta_{\text{Nj}i}$ to make their heights the same.
However, it is clear that every jackknife sample predicts exactly the same $\Delta u_P$, due to the non-zero size $\de = 0.01V$ of the small energy intervals.
We therefore use \de itself to set the uncertainty on the plaquette jump: $\epsilon_{\Delta u_P} = \frac{\de}{6V} \approx 0.0017$.

\begin{figure}[btp]
  \includegraphics[width=0.475\linewidth]{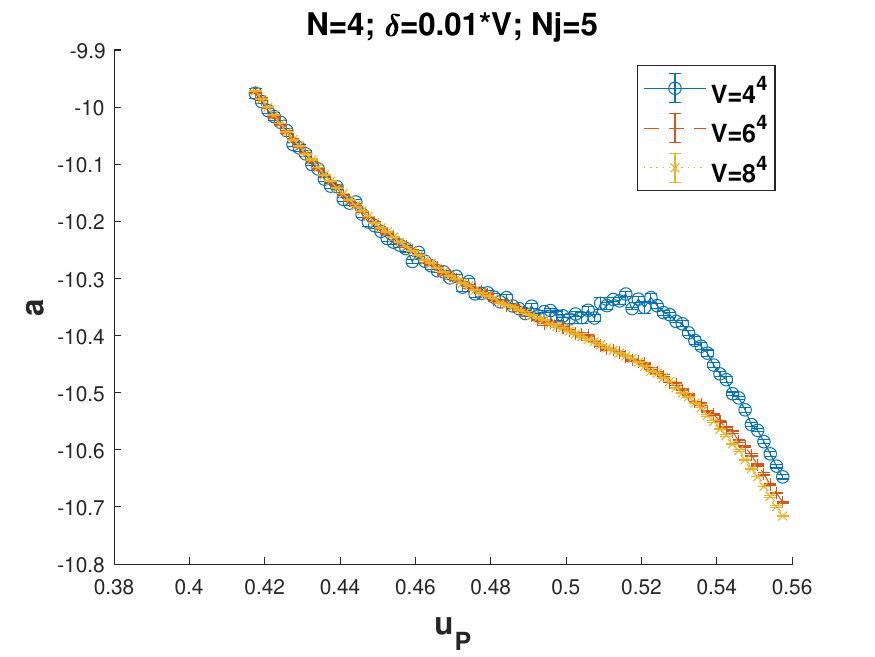}\hfill
  \includegraphics[width=0.475\linewidth]{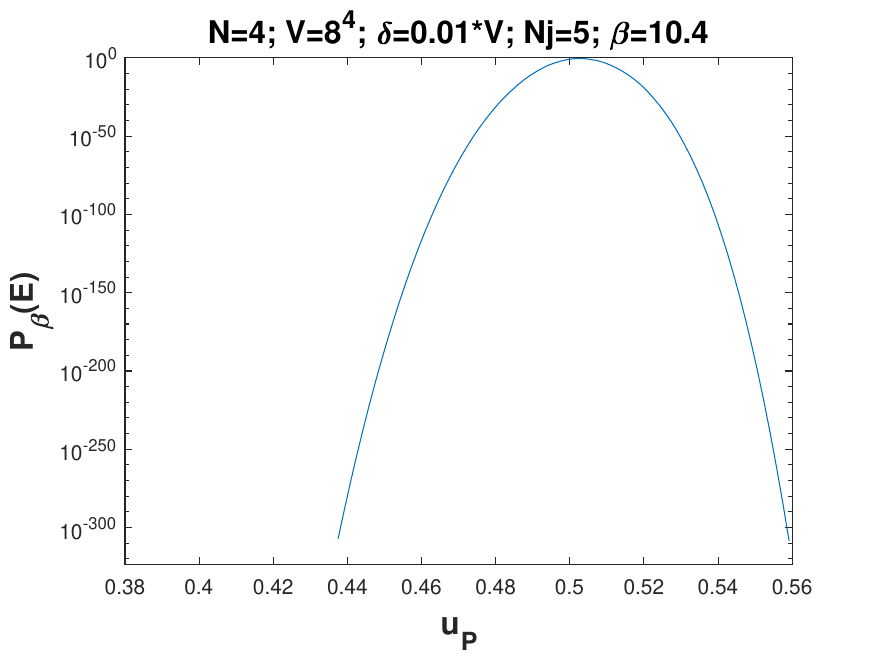}
  \caption{SU(4) results for $V=4^4$, $6^4$ and $8^4$ with an energy interval size of $\de = 0.01V$.  \textbf{Left:} The LLR parameter $a$, with statistical uncertainties obtained by running $\Nj = 5$ independent calculations per interval. \textbf{Right:} The resulting probability density $P_{\beta}$ (omitting uncertainties) for $V=8^4$ at $\beta=10.4$. The single-peak structure persists for all values of $\beta$, confirming the expected continuous crossover.}
  \label{fig:avse_nc4}
\end{figure}

Now that we have reviewed our setup and uncertainty analyses, we present our results for the SU($N$) bulk transition for the action \eq{eq:action}, starting with $N = 4$.
Figure~\ref{fig:avse_nc4} presents our SU(4) results for both $a(u_P)$ (left) and the resulting probability density $P_{\beta}(u_P)$ for $\beta = 10.4$ (right).
As discussed in \secref{sec-LLR}, the two-peak structure in $P_{\beta}(u_P)$ that characterizes a first-order transition corresponds to $a(u_P)$ results that are non-monotonic in the energy.
The $6^4$ and $8^4$ results for $a(u_P)$ in \fig{fig:avse_nc4} are clearly monotonic, and we correspondingly observe a single peak in $P_{\beta}(u_P)$ for all values of $\beta$.
This confirms that the SU(4) bulk `transition' for the action \eq{eq:action} is a continuous crossover, in agreement with \refcite{Lucini:2005vg}.
Note that the smallest $4^4$ volume produces a clearly non-monotonic $a(u_P)$ --- a spurious sign of a first-order phase transition, which leads us to conclude that the $4^4$ volume is too far from the thermodynamic limit to be reliable.

\begin{figure}[btp]
  \includegraphics[width=0.475\linewidth]{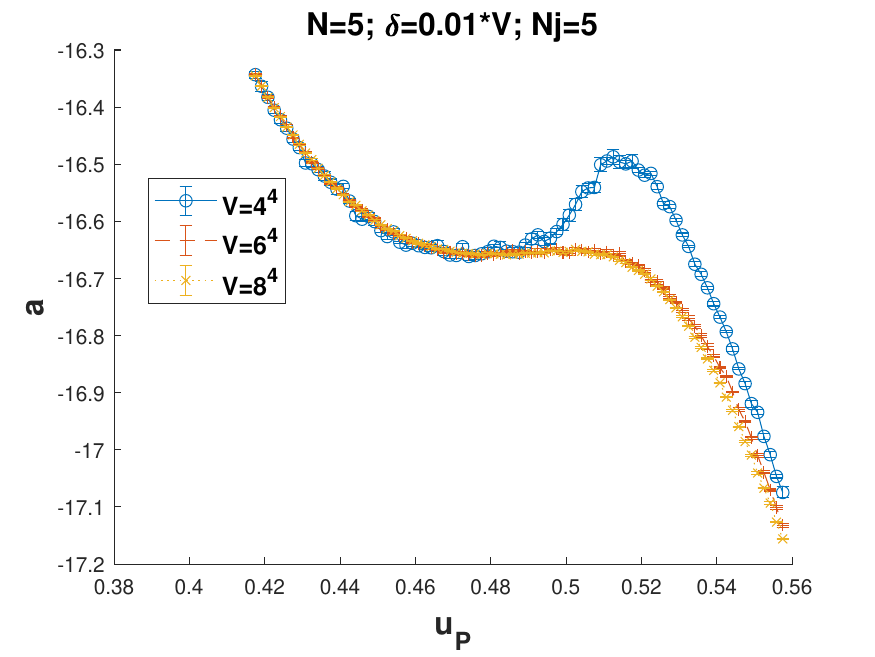}\hfill
  \includegraphics[width=0.475\linewidth]{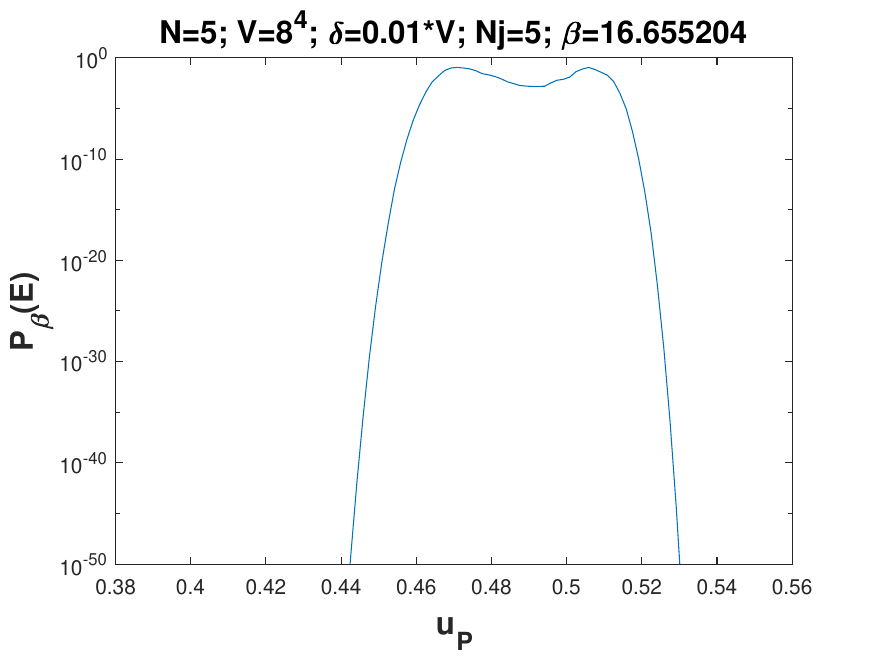}
  \caption{SU(5) results, as in \fig{fig:avse_nc4}.  Here the double-peak structure in $P_{\beta}(u_P)$ signals a first-order phase transition, for which we determine the critical $\beta_{\text{bulk}} = 16.6552(4)$ and $\Delta u_P = 0.035(2)$.}
  \label{fig:avse_nc5}
\end{figure}

SU(5) is the first case for which we observe a true first-order bulk phase transition, in \fig{fig:avse_nc5}.
While it is not obvious by eye, $a(u_P)$ remains slightly non-monotonic for the larger lattice volumes $6^4$ and $8^4$.
It is much easier to see the resulting double-peak structure in the probability density $P_{\beta}(u_P)$ at $\beta_{\text{bulk}} = 16.6552(4)$.
This confirms a first-order transition, for which we can read off the plaquette jump $\Delta u_P = 0.035(2)$ directly from the $P_{\beta}(u_P)$ plot.

\begin{figure}[btp]
  \includegraphics[width=0.475\linewidth]{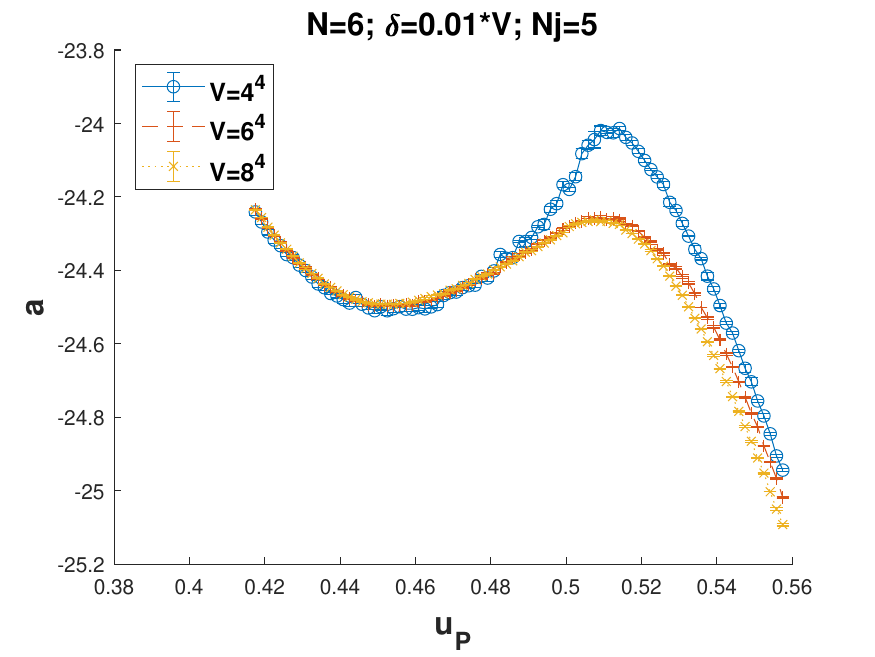}\hfill
  \includegraphics[width=0.475\linewidth]{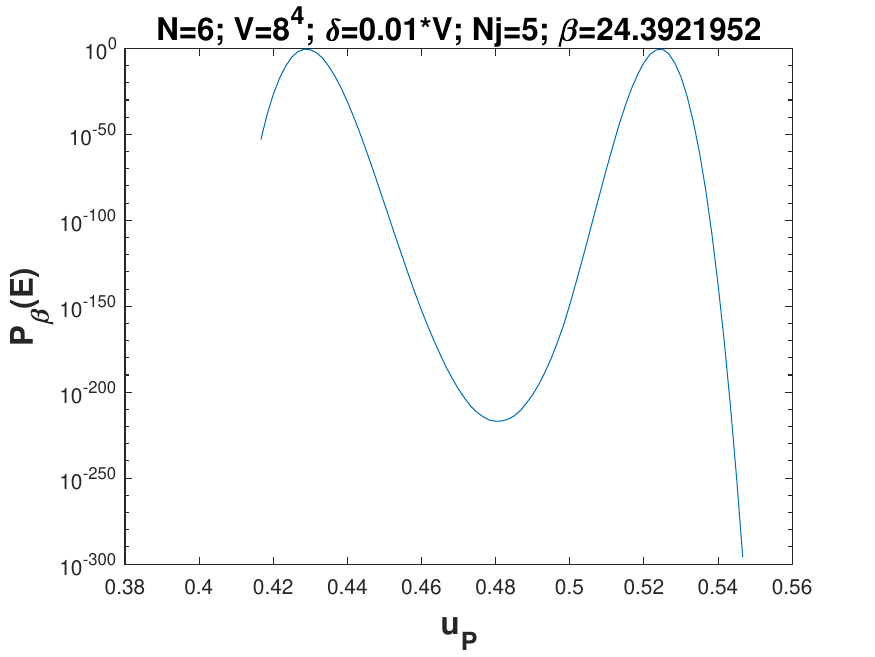}
  \caption{SU(6) results, as in \fig{fig:avse_nc4}, now predicting a first-order transition at $\beta_{\text{bulk}} = 24.39220(26)$ with $\Delta u_P = 0.097(2)$.}
  \label{fig:avse_nc6}
\end{figure}

\begin{figure}[btp]
  \includegraphics[width=0.475\linewidth]{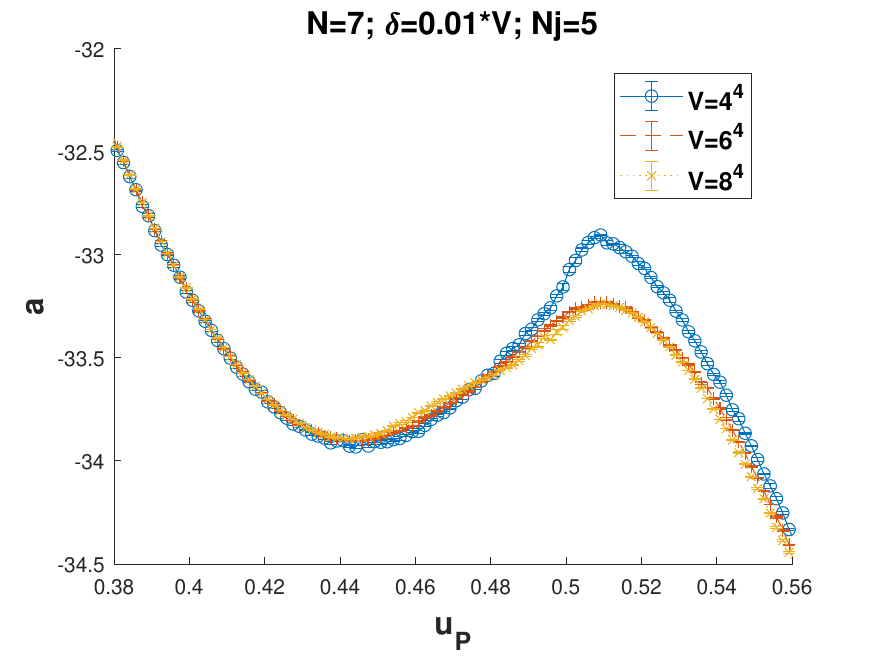}\hfill
  \includegraphics[width=0.475\linewidth]{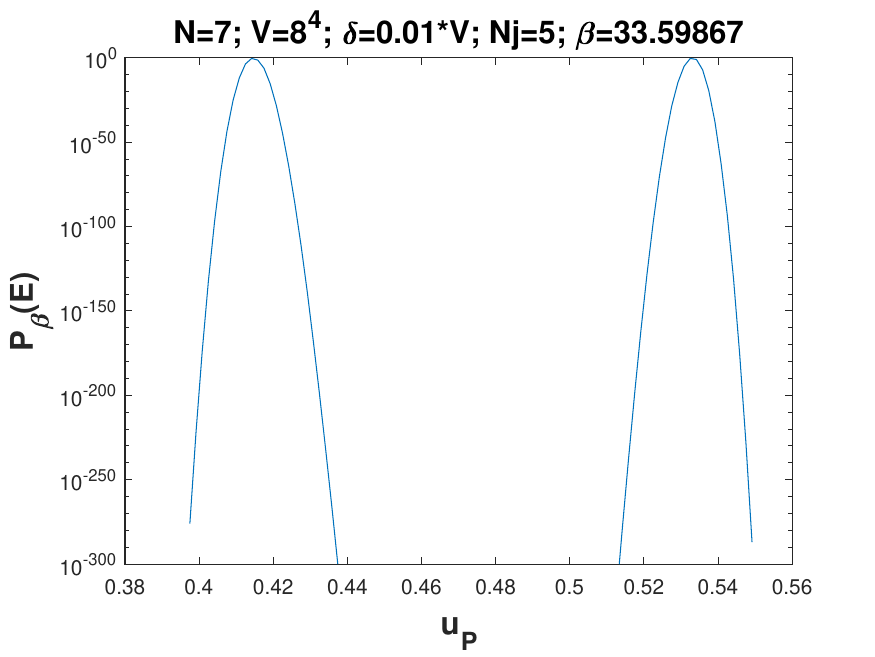}
  \caption{SU(7) results, as in \fig{fig:avse_nc4}, now predicting a first-order transition at $\beta_{\text{bulk}} = 33.59867(28)$ with $\Delta u_P = 0.118(2)$.}
  \label{fig:avse_nc7}
\end{figure}

\begin{figure}[btp]
  \includegraphics[width=0.475\linewidth]{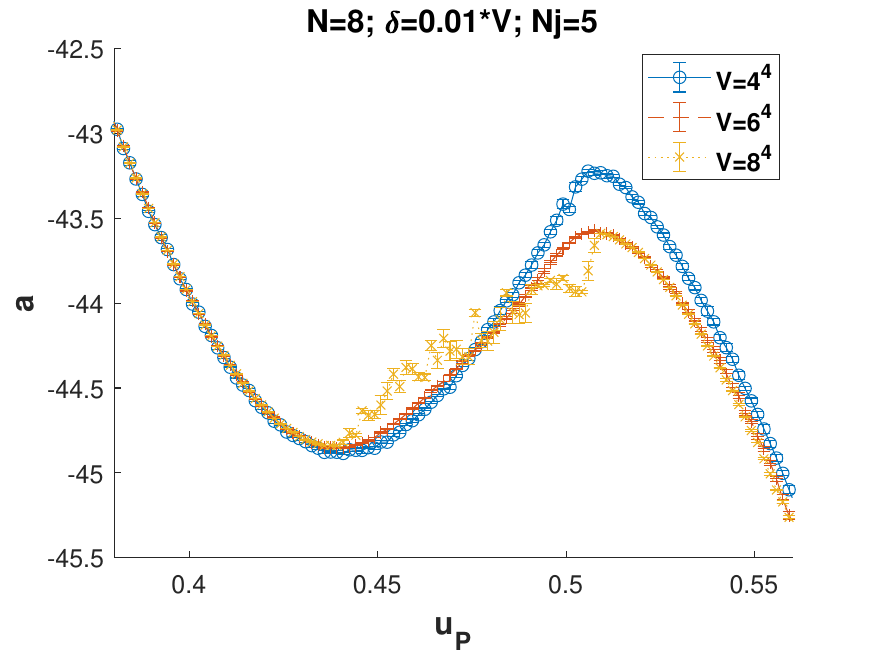}\hfill
  \includegraphics[width=0.475\linewidth]{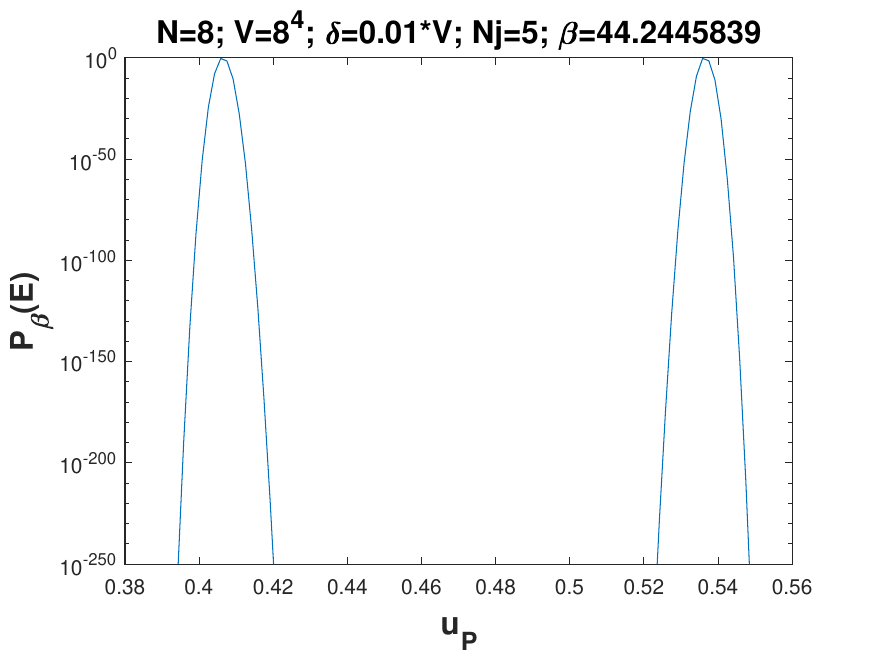}
  \caption{SU(8) results, as in \fig{fig:avse_nc4}, now predicting a first-order transition at $\beta_{\text{bulk}} = 44.2446(32)$ with $\Delta u_P = 0.130(2)$.}
  \label{fig:avse_nc8}
\end{figure}

\begin{figure}[btp]
  \centering
  \includegraphics[width=0.475\linewidth]{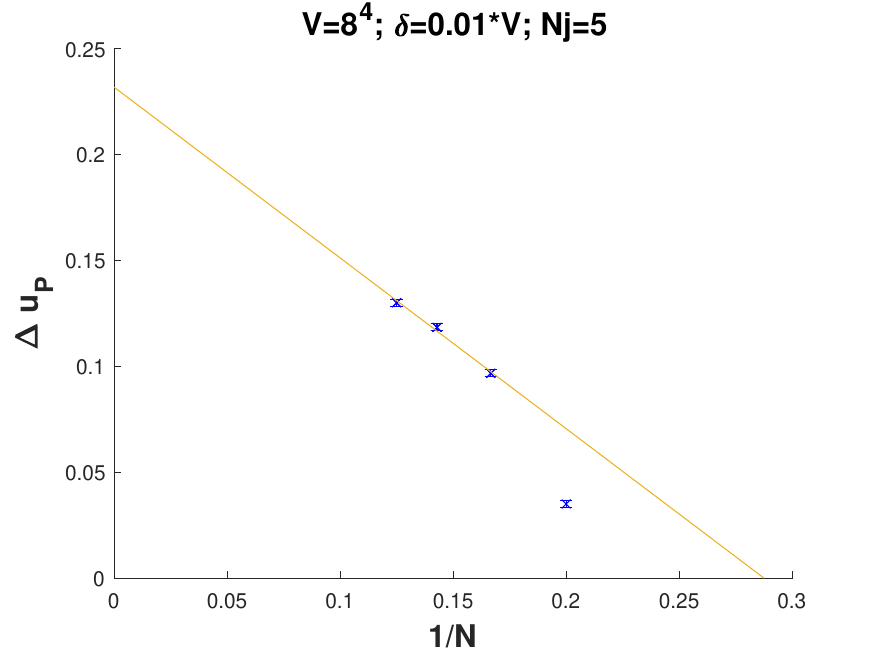}
  \caption{$N$ dependence of the plaquette jump $\Delta u_P$, measured on $V=8^4$ lattices with an energy interval size of $\de = 0.01V$.  A linear fit to the $6 \leq N \leq 8$ results produces $\Delta u_P = 0.230(8) - \frac{0.80(6)}{N}$.}
  \label{fig:upvsN2}
\end{figure}

In the larger-$N$ cases SU(6), SU(7) and SU(8), $a(u_P)$ is clearly non-monotonic for all three volumes, as shown in Figs.~\ref{fig:avse_nc6}, \ref{fig:avse_nc7} and \ref{fig:avse_nc8}, respectively.
As expected, the $\Delta u_P = 0.097(2)$, $0.118(2)$ and $0.130(2)$ at the corresponding first-order phase transitions are all significantly larger than for SU(5).
In particular, in \fig{fig:upvsN2} we are able to fit these three results to the straight line $\Delta u_P = 0.230(8) - \frac{0.80(6)}{N}$, with $\chi^2/\text{d.o.f.} = 1.094/1$ corresponding to $p = 0.296$.
The SU(5) point falls well below this line, confirming that the SU(5) bulk transition for the action \eq{eq:action} is only weakly first order~\cite{Lucini:2005vg}.

Returning to the left panel of \fig{fig:avse_nc8}, let us comment on the strange behavior of the SU(8) $8^4$ results for $a(u_P)$ in the non-monotonic transition region $0.44 \lesssim u_P \lesssim 0.51$.
In this region there is a sudden onset of fluctuations significantly larger than the statistical uncertainties, which we do not observe for any other data set.
This is likely responsible for the $\sim 10\times$ larger uncertainty on $\beta_{\text{bulk}}$ for SU(8) compared to SU(6) and SU(7).
The behavior of $a_j$ in our NR and RM iterations does not resemble what we see when the energy interval sizes \de is made too large for $\log \rho$ to be approximately piecewise linear, and we also obtain fluctuating results for smaller $\de$. 
We similarly obtain fluctuating results if we increase the number of HMC trajectories per iteration, if we increase the number of RM iterations following the 30 initial NR iterations, and if we increase the number \Nj of independent calculations.
Finally, we reran these SU(8) $8^4$ calculations using all three of the other restricted importance sampling algorithms discussed in \secref{sec-sample} --- over-relaxed QHB, full-SU($N$) over-relaxation, and naive MRRTT updates --- in each case imposing hard energy cut-offs. 
In all cases we observed large fluctuations in the transition region, which we will continue to investigate in future work.

\section{Conclusion}
\label{sec-conclusion}
In this work we have applied the LLR density of states algorithm to investigate first-order bulk transitions in pure-gauge SU($N$) lattice Yang--Mills theories with the action \eq{eq:action}.
We have provided a comprehensive review of the algorithm, which allows us to evade the super-critical slowing down of importance-sampling techniques at such first-order transitions.
We focused in particular on algorithmic considerations for calculations with large $N \geq 4$, comparing several restricted importance sampling algorithms used within each stochastic root-finding iteration, and adopting the HMC algorithm to obtain the results presented above.
These results allowed us to confirm~\cite{Lucini:2005vg} that the action \eq{eq:action} features a bulk crossover for $N = 4$, which becomes weakly first-order for $N = 5$ and robustly first-order for $N \geq 6$.

From our results in Figs.~\ref{fig:avse_nc4}--\ref{fig:avse_nc8}, we can appreciate that first-order transitions are easiest to observe with the LLR algorithm when the non-monotonic region of $a(E)$ is large compared to the small energy interval size $\de$, and when the non-monotonicity itself is large compared to the uncertainties on $a(E)$.
Note that reducing \de leads to larger fluctuations in $a_j$ during NR and RM stochastic root finding (Eqs.~\ref{eq:newraph} and \ref{eq:robmon}), increasing the uncertainties on $a(E)$.
This leads us to the conclusion that the LLR algorithm performs best when analyzing strong first-order transitions with large latent heat, for which both of these conditions are easiest to satisfy with relatively large \de and relatively large statistical uncertainties.
This may be counter-intuitive, because it is precisely the opposite behavior to that of more familiar importance-sampling approaches.
The strange fluctuations in the SU(8) $8^4$ results in \fig{fig:avse_nc8}, which we discussed above and continue to study, may raise a caveat to this conclusion.
These fluctuations appear only for the strongest transition we have studied so far, and may be a warning sign that generic difficulties might arise in LLR analyses of even stronger phase transitions.

These considerations also highlight the challenges facing our ongoing investigations of the deconfinement transition for SU(4) Yang--Mills theory, motivated by the Stealth Dark Matter model~\cite{Appelquist:2015yfa, Appelquist:2015zfa, LatticeStrongDynamics:2020jwi} and ongoing observational searches for stochastic backgrounds of gravitational waves~\cite{NANOGrav:2023gor, EPTA:2023fyk, Reardon:2023gzh, Xu:2023wog, Caprini:2015zlo, Caprini:2019egz, Kawamura:2020pcg, AEDGE:2019nxb}.
The deconfinement transition is much weaker than the bulk transition, and the relatively large $30^3\times 6$ lattice volumes we are focusing on translate the already-small latent heat $L_h$ to an even smaller plaquette jump $\Delta u_P \propto \frac{L_h}{N_t^4}$ due to \eq{eq:latentheat}.
This implies a very small non-monotonicity in $a(E)$, but resolving the transition with the LLR algorithm to determine the latent heat and surface tension does appear to be within reach.

In parallel with these studies of the deconfinement transition, it would be both interesting and straightforward to use the LLR algorithm to efficiently map out the bulk phase structure of the action \eq{eq:fundadj} in the fundamental--adjoint ($\beta$--$\beta_A$) plane.
By repeating the work reported here for non-zero values of $r = \frac{\beta_A}{\beta}$, we could locate the first-order bulk transition line in this plane for a sequence of SU($N$) gauge groups, and determine how its critical endpoint moves as a function of $N$.
This would build on earlier work including Refs.~\cite{Creutz:1987xi, Lucini:2005vg, Lucini:2013wsa}, which previously established that the critical endpoint crosses the $\beta_A = 0$ fundamental axis for $N = 5$.
We are also looking forward to applying the LLR algorithm to analyze phase transitions in a variety of other theories, including bosonic matrix models.

\vspace{20 pt}
\noindent \textsc{Acknowledgments:}~We thank Kurt Langfeld, Paul Rakow, David Mason, James Roscoe, Johann Ostmeyer and George Fleming for helpful conversations about the LLR algorithm and related topics. 
Numerical calculations were carried out at the University of Liverpool.
DS was supported by UK Research and Innovation Future Leader Fellowship {MR/S015418/1} \& {MR/X015157/1} and STFC grants {ST/T000988/1} \& {ST/X000699/1}. \\[8 pt]

\noindent \textbf{Data Availability Statement:} The raw data used in this work can be obtained by contacting DS.

\raggedright
\bibliography{main}
\end{document}